\documentclass[a4paper, notitlepage]{article}

\usepackage[a4paper, total={6in, 8in}]{geometry}
\usepackage{graphicx}
\usepackage{grffile}
\usepackage[table]{xcolor}
\usepackage[utf8]{inputenc}
\usepackage{amsmath}
\usepackage{wrapfig}
\usepackage[comma]{natbib}

\renewcommand\harvardyearright[1]{.}

\newcommand{\SubItem}[1]{
    {\setlength\itemindent{15pt} \item[-] #1}
}

\title{Market Impact of Small Orders}
\date{9 January 2022}
\author{Oleh Danyliv \\USoft HTI Inc}

\begin{document}

\maketitle
\begin{abstract}
The article is an empirical study of market impact through order book events. It describes a mechanism of extracting an average participation rate and a market impact of small orders which represent individual slices of large metaorders. The study is based on tick data for futures contracts. It is shown that the impact could be either linear or a concave function as a function of trading volume, depending on the instrument. After normalisation, this dependency is shown to be very similar for a wide range of instruments. 

A simple yet effective model for market impact estimation is proposed. This model is linear in nature and is derived based on straightforward microstructure reasoning. The estimation shows satisfactory results for both concave and linear market impact volume dependencies. 

\end{abstract}


\section{Introduction}

A large amount of empirical results has been accumulated over the
years, concerning the relation between the size of the trading order and the
resulting average price change. These empirical results come from the analysis of the executions of large customer orders by brokerage firms or exchanges (\citealp{Toth2011}; \citealp{Ferraris2008}; \citealp{Kissel2006}; \citealp{Moro2009}). Empirical results usually show a linear volatility dependence of the impact and concave (square root type)  volume dependency.

The theoretical approach to understanding market impact was proposed by \cite{Kyle1985}, where market participants were split into three kind of traders: a single risk neutral {\em informed trader}, random {\it noise traders} and competitive risk neutral {\it market makers}. It is shown that the optimal strategy for the {\it informed trader} consists of breaking its metaorder into pieces and executing them incrementally at a uniform rate,
gradually incorporating its information into the price. The price will then grow linearly with volume and the linear coefficient is called {\it Kyle's } $\lambda$. The linearity of the market impact as a function of volume is in accordance with \cite{Huberman2004}, who showed that if the price impact is permanent, only linear price impact functions rule out arbitrage, i.e. the incremental impact per share remains constant during the metaorder life.

\cite{AlmgrenChriss2000} introduced a decomposition of the total impact into
a permanent market impact, which corresponds to changes in the equilibrium price due to trading, and a temporary impact, which refers to temporary imbalances in supply in demand caused by trading, leading to temporary price movements away from equilibrium. This decomposition is used by most practitioners in execution models. \cite{Almgren2005} fitted the parameters of this model using a large data set from the
Citigroup US equity trading desks: the permanent market impact was found to be a linear function of volume and the temporary market impact follows 3/5 power law, which is close to the square root dependency.

The analysis of market impact based on order book events (limit orders, market orders and cancellations) for US stocks was conducted by \cite{ContStoikov14} and the results were confirmed more recently by \cite{BugaenkoPhD}. The authors had found a linear relationship  between {\it order flow imbalance} (which reflects a change in order book with time) and price changes. The slope of this relationship was found to be inversely proportional to the volume in the order book. \cite{ContStoikov14} also studied the impact dependency on {\it trade imbalance}, but that dependency was found to be noisy and less robust than the order flow imbalance.

In this paper the study of the market impact of small orders will be based on the order book events for futures markets. Small orders correspond to the impact of slices of a large metaorder. They are easier to analyse since the execution time is usually small and one can assume that there is only one {\it informed trader} present on the market.

\section{Trade imbalance} \label{sec:Model}

\begin{figure} [h]
    \centering
\includegraphics[width=0.65\textwidth]{"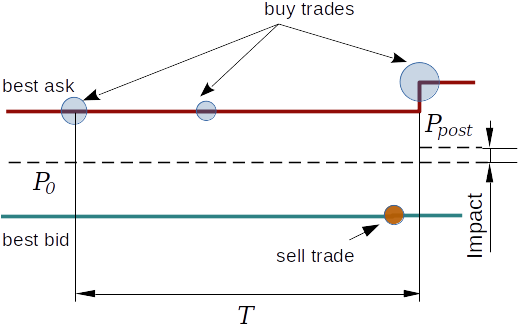"}
    \caption{The formation of the order trade imbalance during time $T$. It starts from a buy trade on the best offer level. It continues until $V_I \geq V_T$, when he influence of a new trader willing to buy $V_T$ contracts/shares stops and market returns to the state of equilibrium.}
    \label{fig:order_flow_imbalance}
\end{figure}

Market impact can be defined as the difference between the post trade mid-price $P_{post}$, which is based on the first quote following the trade and the mid-price before the trade $P_0$ (see Fig.\ref{fig:order_flow_imbalance}):

\begin{equation}
I = P_{post} - P_{0}, \label{I_definition}
\end{equation}

It is important to distinguish an order slippage from the market impact. It is clearly visible on an execution of an order with minimal volume. An aggressive order with a minimal volume will have to pay up the spread, which will define its slippage. The market impact (\ref{I_definition}) of such an order is likely to be zero since the order book usually has enough limit orders to provide the liquidity for a minimal trade without exhausting best bid/offer level. Therefore, the volume dependence of the market impact with this definition should start at zero and will be close to zero value for small orders. 

Let's consider a market in a perfect equilibrium state: the buy volume equals the sell volume during a prolonged period of time; the flows of new market participants on both sides of the market are equal and balance each other. Using terminology of \cite{Kyle1985}, the market consists only of {\it noise traders} and {\it market makers}. If there are no breaking news on the market, the price is likely to stay the same at the fair price level $P_{post}$ or slightly fluctuate around this value. The balance is then disturbed by an {\it informed trader} who is willing to purchase $V_T$ contracts - the order, relatively small compared to the total order book. There are two main scenarios of the execution:

a) To minimise the timing risk, a new buyer initiates an aggressive trade. Then we will observe an immediate buy trade at the price of the best offer.

b) The buyer is willing to take the risk and executes the order passively.
Then one should consider the fact that there is always an urge to execute a small order on any market. Most of the market participants are bounded by an execution schedule which is dictated by trading strategy: in the case of 
the time-weighted average price (TWAP) strategy a trading algo has to execute an equal number of shares for every slice time $T$. One of the most popular equity strategies, volume-weighted average price (VWAP), has to split an order to follow the volume profile curve. For a percent of volume (PoV) order, a trading algo has to trade a specific fraction of volume traded on the market, implementation shortfall (IS) strategies are are usually written as PoV and have some volume dependent schedule. Therefore, at some point one of the market participants will need to give up their passive position in the order book and cross the spread to catch up with the explicit or implicit time or volume schedule. As the result, the {\it informed trader} will steal a passive order execution from other participants. 

In both cases an aggressive trade is observed on the opposite side of the order book. In general, not one but $N$ trades with signed volume $\varepsilon_i V_i$ will be observed until the equilibrium is restored and the number of buyers will match the number of sellers again. The sum of signed volumes is called {\it trade imbalance} and is formally defined as:

\begin{eqnarray}
V_I & = & \sum_{i=1}^N \left(\varepsilon_i V_i \right), \label{V_I} \\
\varepsilon_i & = & 
\left\{ 
  \begin{array}{ c l }
    1 & \quad \textrm{if } P_i \geq P_{ASK}, \\
    -1 & \quad \textrm{if } P_i \leq P_{BID}, \\
    0  & \quad \textrm{otherwise}. \nonumber
  \end{array}
\right.
\end{eqnarray}

$V_I$ will be equal exactly to trading volume $V_T$ in the case a) of an aggressive buyer or it will be close to this volume in the case b) of a passive execution. That is why one can use {\it trade imbalance} as a proxy of the volume traded by the {\it informed trader}.

The following algorithm, illustrated in Fig. \ref{fig:order_flow_imbalance}, will be used further to examine the dependency of market impact on trading volume:
\begin{description}
 \item[Start:] We start from the assumption that the instrument is in the state of the equilibrium at price $P_{0}$ and $V_I=0$.
 \item[Calculation:] Signed trades are summed up according to (\ref{V_I}). 
 \begin{itemize}
     \item If at some point $V_I$ crosses zero without reaching value $V_T$, that means that we are still in the equilibrium phase and $P_{0}$ is reset to the current mid-price.
      \item When the volume imbalance becomes larger or equal to the trade volume $V_T$, then the volume count is  stopped. 
 \SubItem {If $\frac{V_I-V_T}{V_T} \ll 1$, the market impact (\ref{I_definition}) is calculated;}
 \SubItem {If $\frac{V_I-V_T}{V_T} \sim 1$, then {\ trade imbalance} cannot be used as a proxy of the trade volume and the result is ignored.}
      \item Begin new trade simulation from {\bf Start} using succeeding order book events.
\end{itemize}

\end{description}

The calculating procedure starts at the beginning of the trading session and cannot go beyond the end of the session. Since only $V_I \simeq V_T$ values are selected, the {\it trade imbalance} dependency will be equivalent to a trading volume dependency.

The methodology described will perfectly describe market impact in the ideal case, which is never realised in real life. First of all, it assumes that there is only one {\it informed trader} on the market, who accumulates all trades which permanently move the price. It aggregates all such traders in one collective {\it informed trader}. Additionally, it assumes one exchange: if the liquidity escapes to another trading venue, the value of the impact will change.  

\section{Market impact on the example of Crude Oil Futures} \label{sec:calculations}

Futures markets are the most suitable to provide research of the impact (\ref{I_definition}) because they primarily trade on centralised venues. Futures trading is less fragmented, compared to, for example, European equity markets where orders can trade in dark pools, multilateral trading facilities (MTFs), systematic internalisers (SIs), on auctions or off the counter (OTC)\footnote{To monitor the volume of trades on different venues, a trading platform provider Fidessa PLC developed comprehensive {\it Fidessa Fragmentation Index} which was published in marketing purposes.}. Comprehensive equity execution algorithms send orders simultaneously to multiple exchanges to maximise liquidity and provide {\em best execution}. 

We will test the influence of $V_I$ using different futures contracts in a two week period using {\it Refinitiv} Level 1 \footnote{Level 1 tick data provides execution events and quote changes on the best bid and best offer levels.} data in the time interval from 07/09/2021 till 16/09/2021. Despite a short period of time, the amount of data is sufficient to obtain statistically significant results. For example, for the Crude Oil contract {\it CLc1} there were around 80 million Level 1 data events, 1.34 million of which where trades (matching events). We will follow the data provider notation, the nearest delivery month is marked as {\it c1}, the next delivery month is {\it c2}, etc \footnote{The study will be focused on the most liquid delivery month which is usually September, but for some instrument like Gold, the most traded month is {\it c4} which is December. }. 

\begin{figure} [h]
	\begin{minipage}[b]{0.48\textwidth}
  		\centering
   		\includegraphics[width=\textwidth]{"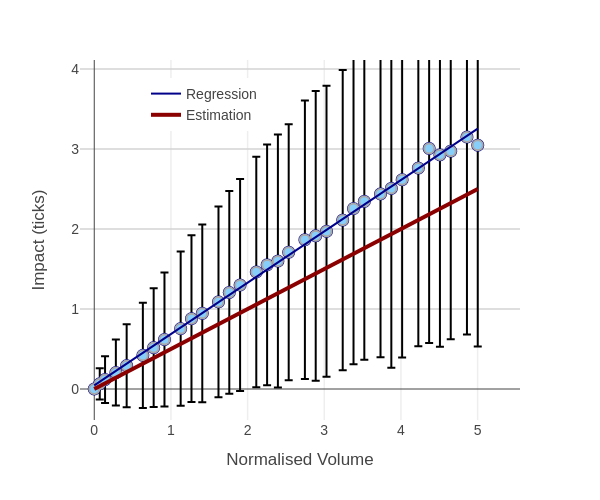"}
    	\caption{The average market impact for {\it CLc1}. Error bars are equal to one standard deviation of impact distributions, assuming a Gaussian error model. The red line is the estimation given by formula (\ref{I_E}).}
    	\label{fig:oil_simulations}
	\end{minipage}
	\quad    
	\begin{minipage}[b]{0.48\textwidth}
    	\centering
		\includegraphics[width=\textwidth]{"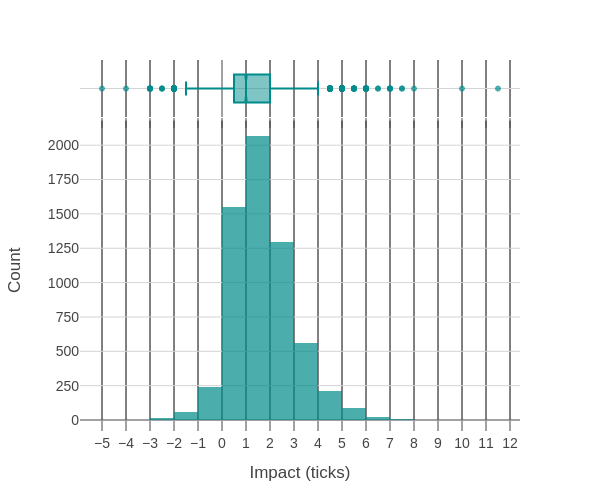"}
    	\caption{The distribution of market impacts (\ref{I_definition}) for {\it CLc1} when the order size equals two average touch sizes (28 contracts). Above the histogram the box diagram shows the first quartile, median, third quartile, maximum and outliers,}
    	\label{fig:histogram}
	\end{minipage}
\end{figure}

\begin{figure} [h]
    \centering
\includegraphics[width=0.5\textwidth]{"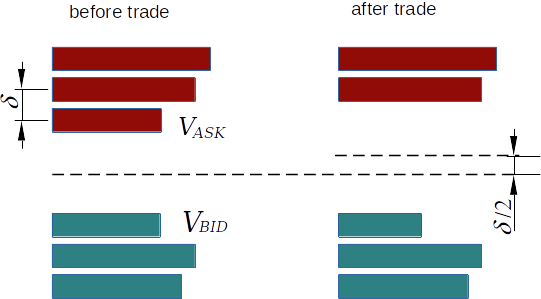"}
    \caption{The aggressive order of the size, equal to the volume on best offer, will have an impact $\delta/2$.}
    \label{fig:aggressive_order}
\end{figure}

The result of the impact calculations for September 2021 Crude Oil contract {\it CLc1} is shown on Fig.\ref{fig:oil_simulations}. The impact is measured in ticks which will be denoted by $\delta$ - one tick is the instrument dependent minimum difference between price levels which is defined by the exchange (see Fig.\ref{fig:aggressive_order}); for the Crude Oil contracts $\delta = \$0.01$. The volume on the chart is normalised:  it is divided by the average touch size. In this study the price drift is ignored and therefore the most appropriate choice is a symmetrical average touch size which treats buy
and sell sides of the market equally :
\begin{equation}
\left\langle V_{Touch} \right\rangle_t = \frac{\left\langle V_{BID} \right\rangle_t + \left\langle V_{ASK} \right\rangle_t}{2},
\label{V_touch}
\end{equation}
where $\left\langle V_{BID} \right\rangle_t$ is a time weighted volume on the best bid price level. For the reference, $\left\langle V_{Touch} \right\rangle_t \simeq 14$ contracts for September 2021 Crude Oil futures. 

Fig. \ref{fig:oil_simulations} demonstrates that, despite the fact that each individual trade has a randomly distributed impact, the  average impact is a linear function of trading volume with very high coefficient determination $R^2=0.996$.  
Error bars represent one standard deviation of the impact distributions at the current trade volume. They show that the error of the impact calculation is quite high. The distribution itself is shown on Fig. \ref{fig:histogram},  the histogram of impacts at a given volume. Impacts could take only discrete tick values, forming a natural histogram. The distribution is not symmetrical with the longer tail for positive values. The impact of an individual trade can be negative when {\it noise traders} temporarily shift market price in the opposite direction to the trade direction.

The linear regression of the market impact will be denoted as:
\begin{equation}
I = \mu_I + \lambda_I v,
\label{linear_impact}
\end{equation}
where the $\mu_I$ is the intercept and the slope $\lambda_I$ is {\it Kyle's lambda} which provides the relation between the price change and the normalised volume $v$: 
\[
v = \frac{V_T}{\left\langle V_{Touch} \right\rangle_t}.
\]

There is an easy way to make a rough estimation of the slope of this dependency in the idealistic situation when the volume on touch stays the same during all trading sessions. If the aggressive order takes the whole liquidity on a best bid/offer level, then this level will step up by value $\delta$, creating an estimated impact $I_E(v = 1) = \delta/2$ (see Fig.\ref{fig:aggressive_order}). 
The impact of a zero volume is zero, therefore, for arbitrary trading volume,

\begin{equation}
I_E = \frac{\delta}{2} v,
\label{I_E}
\end{equation}
giving value for the expected slope $\lambda_E = {\delta}/{2}$. The dependency (\ref{I_E}) is shown on Fig.\ref{fig:oil_simulations} alongside the trend line. Despite a strong assumption of constant touch volume, this formula gives a good initial expected value for market impact: for {\it CLc1} contracts, it underestimates the real value by only 28\%, with even smaller errors for some other instruments (Table \ref{tab:table1}). 

The formula for estimated impact $I_E$ does not include volatility dependency, which was reported in all empirical studies (see for example by \cite{Almgren2005}). This result is expected: formulas (\ref{linear_impact}, \ref{I_E}) describe market impact of a fraction (a slice) of large metaorder. The next portion of the order will be released to the market according to the execution schedule of the algorithm after some period of time. During this time, the price of  the instrument will change under influence of other {\it informed traders}, reaching a new point of equilibrium, and the total impact of the metaorder will incorporate volatility.

The algorithms described in Section \ref{sec:Model} allows the extraction of an average value of impact. This result takes into account all types of executions. But it is a mixed bag: some market participants try to avoid risk and trade aggressively, some try to capture the spread. To understand meaning of market average, one needs to evaluate additionally an order execution time and the participation rate. Both these measures could be calculated alongside the impact. 

\begin{figure} [h]
	\begin{minipage}[t]{0.48\textwidth}
  		\centering
		\includegraphics[width=\textwidth]{"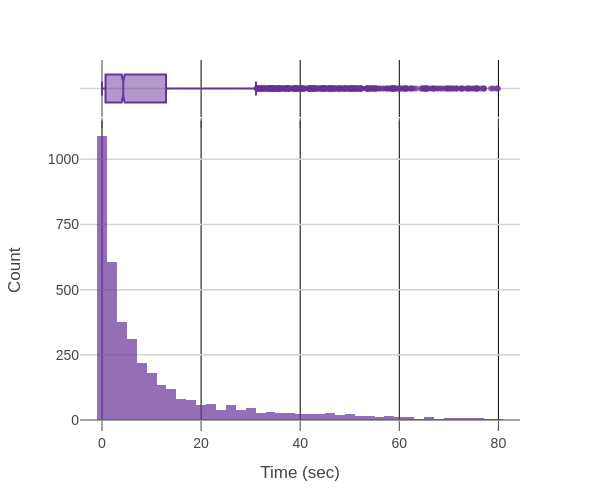"}
    	\caption{The execution time of two average touch sizes of Crude Oil futures expiring in September 2021. The box plot is shown above the histogram.}
    	\label{fig:time_distribution}
	\end{minipage}
	\quad    
	\begin{minipage}[t]{0.48\textwidth}
    	\centering
    	\includegraphics[width=\textwidth, height=0.9\textwidth]{"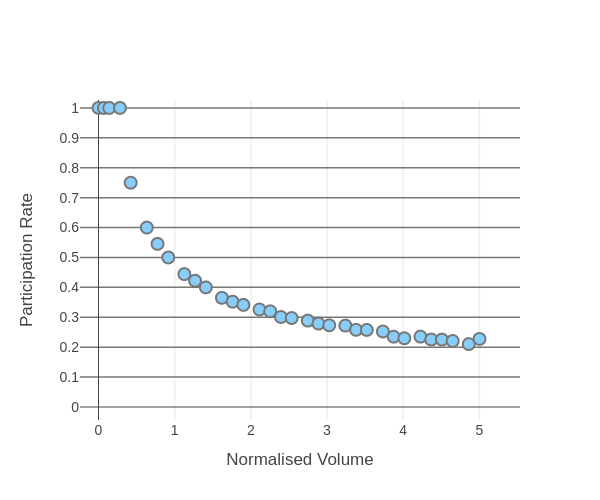"}
    	\caption{The median participation rate as a function of volume for CLc1.}
    	\label{fig:participation_rate}
	\end{minipage}
\end{figure}

Fig. \ref{fig:time_distribution} shows the distribution of the execution time for {\it CLc1} futures contracts at the fixed normalised volume $v = 2$ (or $V_T \simeq 28$ contracts). Values vary in a large range from zero seconds to a few hours. 2.5\% of all orders were purely aggressive with the zero execution time. But the average execution time was found to be 18 seconds with an even shorter median time of 4.6 seconds. This analysis removes outliers, giving the estimation for Crude Oil average trading rate of {\it informed traders} from 1.55 to 6 contracts per second.  

The average participation rate of {\it informed traders} could be estimated by calculating, alongside volume $V_I$, a volume produced by {\it noise traders}. The ratio of volume $V_I$ to the total volume traded  during time $T$ is a proxy of participation rate. For small orders with $v < 1$ this ratio will overestimate the real value: the majority of small orders will be executed quickly, possibly in one market order, with zero execution time and the participation rate 100\% (Fig. \ref{fig:participation_rate}). But the aggressive execution might have been preceded by unsuccessful passive stage (where an order was not filled) and the total time of the order was not zero. The effect of possible unaccounted passive order stage will decline for orders with larger sizes ($v > 1$) and longer average execution time. The figure shows that the participation rate decays with volume and reaches some asymptotic level. This level is likely to be the real value of the participation rate for the instrument. For orders of Crude Oil futures this value stays around 21\%.

\section{Other instruments} \label{sec:other_instruments}

\begin{figure} [h]
    \centering
\includegraphics[width=0.9\textwidth]{"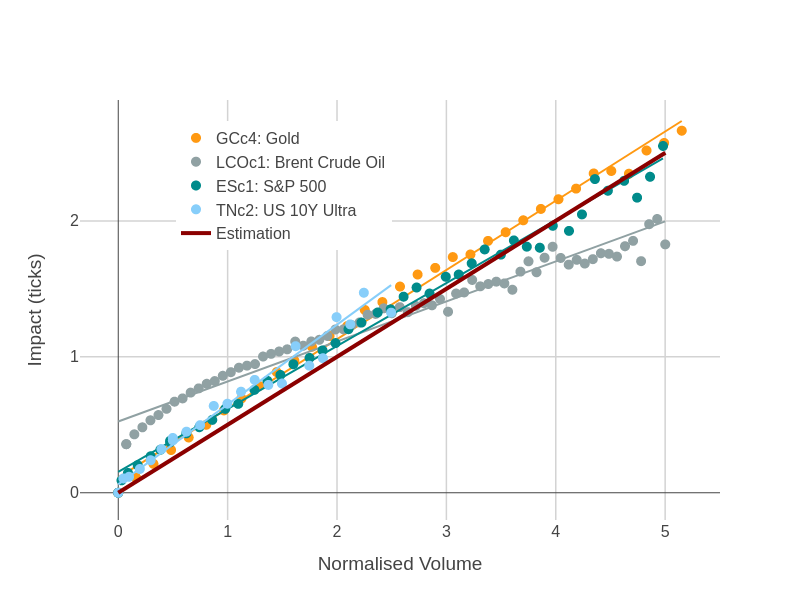"}
    \caption{The average market impact as a function of volume for different instruments.}
    \label{fig:multi_simulations}
\end{figure}

Futures contracts are characterised by expiry dates. Contracts with deliveries far in the future are usually traded infrequently, which makes them less suitable for the analysis. To avoid confusion, the contracts with higher trading volume will be selected for the study. Fig. \ref{fig:multi_simulations} shows market impact $I$ for various unrelated instruments: Gold, ICE Brent Crude Oil, S\&P 500, US Treasuries\footnote{The US 10 Year Ultra Treasuries futures chart is terminated earlier as there was not enough large trade imbalances in the data.}. Most of them show similar linear linear volume dependency. 

The parameters of linear fits for the market impact (\ref{linear_impact}) are shown in Table \ref{tab:table1}. All linear regressions have a very high coefficient of determination $R^2$, which in all cases are not less than 0.94. The result is statistically highly significant: the $p$-value is well below 0.001 significance level, rejecting the null hypothesis of no dependence of market impact from volume. 

The table also provides a comparison of the regression with the estimated result (\ref{I_E}).
The estimation  predicts a zero intercept and value $\lambda_E=\delta/2$ for the slope. $\lambda_{err}$ in the table is the percentage difference of the real and the estimated values. It is clear that formula (\ref{I_E}) is quite approximate and works well only for some instruments like Gold ({\it GCc4}) and Natural Gas ({\it NGc1}). 
In contrast, NYMEX Crude Oil for the next month delivery contract ({\it CLc2}) and the ICE Brent Crude Oil for current month (September) delivery ({\it CLOc1}) have high approximation error $\lambda_{err}$ and non-zero intercept $\mu_I$. 
Fig.\ref{fig:multi_simulations} shows that {\it LCOc1} contract has a complex order volume behaviour: for small volumes $v < \frac{1}{2}$ the volume dependence is concave with the impact very quickly reaching half a tick value even for small volumes, but with larger volumes it could be considered linear, which is confirmed by a high value of the coefficient of determination.

The last column in the table is the estimated participation rate, which is calculated as the median participation rate value for the largest volume obtained during simulations ($v  = 2.5$ for the US Treasuries futures and $v  = 5$ for the rest of the instruments). These values strongly differ between instruments and correspond to an averaging over all trading strategies. They are in a wide range from 13.7\% for bonds futures to 40.3\% for Brent Crude. These numbers are not far from optimal values of participation calculated for equities by \cite{Gueant2012}, who found that the optimal participation rate of bulk orders should be within the range from 17\% to 28\%; empirical results for stocks based on customer data conducted by \cite{Moro2009} showed average participation rate 17\% for Spanish Stock Market and 34\% for London Stock Exchange.\footnote{These numbers will be different for small market participants: hedge funds are likely to trade with smaller 5\%-10\% participation rates.}  

\begin{table}[h!]
  \begin{center}
    \begin{tabular}{|l|c|c|c|c|r|c|c|c|} 
      \hline
       \rule{0pt}{3ex} 
       \rule[-2ex]{0pt}{0pt} 
      \textbf{RIC} & $\left\langle V_{Touch} \right\rangle_t$ &$\delta$& $\mu_I$ & $\lambda_I$& $\lambda_{err}$& $R^2$ & $p$-val& {\it Part.Rate}\\
      \hline \hline
      \rm{CLc1} & 14.2 & \$0.01 & 0.05& 0.64 &  28.0\% & 0.997 &$1.2 \times 10^{-40}$& 21.0\%\\
      \rowcolor{lightgray}
      \rm{CLc2} & 14.4 & \$0.01 & 0.33& 1.10 & 120.0\% & 0.981 & $4.0\times 10^{-30}$ & 16.6\%\\
      \rm{NGc1} &  4.7 & \$0.001& 0.13& 0.52 &   4.0\% & 0.995 & $2.1\times 10^{-29}$ & 36.3\%\\
      \rowcolor{lightgray}
      \rm{LCOc1}& 13.6 & \$0.01 & 0.52& 0.29 &  41.2\% & 0.947 & $4.0\times 10^{-19}$ & 37.0\%\\
      \rm{GCc4} & 6.2 & \$0.1  & 0.11& 0.50 &   1.9\% & 0.994 & $3.1\times 10^{-36}$  & 29.0\%\\
      \rm{ESc1} & 33.7 & \$0.25 & 0.15& 0.46 &   7.4\% & 0.991& $9.3\times 10^{-38}$  & 25.8\%\\
      \rm{WDOc1}& 154.3& \$0.5  & 0.17& 0.37 &  26.9\% &  0.992 & $3.2\times 10^{-33}$  & 40.3\%\\
      \rm{TYc2} & 1888  & \$$\frac{1}{64}$&-0.01&0.66&  31.6\% & 0.950 & $6.8\times 10^{-16}$ & 13.7\%\\
      \rm{TNc2} & 192.4 & \$$\frac{1}{64}$&0.06&0.59&  18.0\% & 0.942 & $3.7\times 10^{-15}$  & 15.8\%\\
      \hline
    \end{tabular}
  \end{center}
    \caption{Linear regression of the market impact for different instruments. Instruments with the concave market impact dependency are shaded.}
    \label{tab:table1}
\end{table}

\section{Conclusions} \label{sec:conclusions}

Market impact of an order depends on many factors: the volume of the order, trader's strategy, volatility, current market conditions, market fragmentation. This complexity seems to be much reduced for futures instruments, which are traded on a central venue. In the case of central order book, {\it trade imbalance} could be interpreted as a trading volume of an {\it informed trader}, which disturbs market equilibrium and affects the price. 
It is shown that the dependency of the market impact on trading volume is either a linear or concave function. Even when this dependency is non-linear, a linear regression has a large coefficient of determination and still can be used. 

A simple, yet effective formula to estimate a market impact of small orders based on the microstructure parameters is proposed. It was tested on different groups of futures instruments: energy, metals, index futures, currency futures and interest rate futures. The approximation works better for some instruments than for others, but still can be used as a rule of thumb for the market impact estimation of a single slice in the algorithmic execution of a large order.

It is shown that the study of market impact using {\it trade imbalance} allows the estimation of medium participation rate for current instrument, which could be used in PoV and IS trading algorithms as a reference point.

\section{Acknowledgements} \label{sec:acknowledgement}
The author thanks Thomas Strange, Yaohui Yu and Tom Middleton for help and valuable comments.

\bibliographystyle{agsm} 
\bibliography{refs} 



\end{document}